\documentclass[12pt]{article}
\usepackage{scicite}
\usepackage{times}
\usepackage{graphicx}
\usepackage{xcolor}

\newenvironment{sciabstract}{%
\begin{quote} \bf}
{\end{quote}}

\newcounter{lastnote}

\title{Imaging the onset of the resonance regime in low-energy NO-He collisions}

\author
{Tim de Jongh$^{1 \ast}$, Matthieu Besemer$^{1 \ast}$, Quan Shuai$^{1 \ast}$, Tijs Karman$^2$,\\
Ad van der Avoird$^1$, Gerrit C. Groenenboom$^{1 \ast \ast}$  Sebastiaan Y.T. van de Meerakker$^1$ $^{ \ast \ast}$ \\  \\
\\
\normalsize{$^1$Radboud University, Institute for Molecules and Materials}\\
\normalsize{Heijendaalseweg 135, 6525 AJ Nijmegen, the Netherlands}\\
\normalsize{$^2$ Institute for theoretical atomic molecular and optical physics}\\
\normalsize{Center For Astrophysics Harvard \& Smithsonian 60 Garden Street, Cambridge, MA 02138, USA}\\
\normalsize{$^{\ast}$ These authors contributed equally to this work}\\
\normalsize{$^{\ast\ast}$To whom correspondence should be addressed;}\\
\normalsize{E-mail: basvdm@science.ru.nl, gerritg@theochem.ru.nl}}

\date{}

\begin{document}

\baselineskip20pt

\maketitle

\begin{sciabstract}

At low energies, the quantum wave-like nature of molecular interactions result in unique scattering behavior, ranging from the universal Wigner laws near zero Kelvin to the occurrence of scattering resonances at higher energies. It has proven challenging to experimentally probe the individual waves underlying these phenomena. We report measurements of state-to-state integral and differential cross sections for inelastic NO-He collisions in the 0.2 - 8.5 cm$^{-1}$ range with 0.02 cm$^{-1}$ resolution. We study the onset of the resonance regime by probing the lowest-lying resonance dominated by \emph{s} and \emph{p} waves only. The highly structured differential cross sections directly reflect the increasing number of contributing waves as the energy is increased. A new NO-He potential calculated at the CCSDT(Q) level was required to reproduce our measurements.
\end{sciabstract}%

\newpage
The study of collisions between atoms and molecules at the full quantum level has been an important research goal in chemistry and physics for decades \cite{Levine:reaction-dynamics}. To date, the internal rotational and vibrational states of molecules have received most attention, and a rich variety of experimental methods has been developed to select a single quantum state before the collision, and to probe the occupied quantum states of the products \cite{Schiffman:IRPC14:1995}. The wealth of experimental and theoretical state-to-state scattering studies has contributed immensely to our present understanding of molecular interactions \cite{Brouard:book:tutorials}. 

Yet, in addition to these internal states, the angular momentum associated with the relative motion of the particles is also quantized. It is described by a set of orbital angular momentum states, or partial waves, with integer quantum number $\ell$ which takes the values $\ell=0,1,2,....$ with names $s$, $p$, $d$, ...-wave, respectively. The contribution of each partial wave to a scattering cross section describes how the reagents transform into products at the most fundamental state-to-state level, and contains all information about the scattering event. 

The influence of a single partial wave to a collision event is in principle directly encoded in the differential cross sections (DCSs), that can be probed experimentally. For atom-atom collisions, each wave will result in a unique angular distribution in which the number of nodes observed in a DCS equals the value of $\ell$ \cite{McDonald:Nature535:122}. That is, $s$-wave scattering leads to an isotropic DCS, $p$-wave scattering results in a single node, etc.. For molecular systems with anisotropic interactions, multiple partial waves contribute and interfere with each other, resulting in a more complicated DCS pattern from which the partial wave composition cannot be directly discerned. The identification of the partial wave composition of a collision event from experimental observations has therefore remained a formidable challenge \cite{Perreault:science358:356}, preventing state-to-state experiments at the full partial wave level. 

The effects of individual partial waves may best be observed in situations where only a very limited number of interfering waves contribute. This number critically depends on the particles' de Broglie wavelengths, and hence the collision energy \cite{Bohn:Science357:1002,Krems:ColdMolecules}. If the energy is too high, a large number of waves contribute, such that the collisions may be regarded as semi-classical and the dynamics can often be described by semi-classical models. There is no hope to disentangle the contribution of individual waves from experimental observations in this regime. By contrast, at energies approaching zero Kelvin, only the $s$-wave contributes leading to universal rules for the energy dependence of collision cross sections in the form of Wigner's threshold laws \cite{Wigner:PhysRev73:1002}. In this often referred to as the \emph{ultracold} regime, the $s$-wave results in differential cross sections that contain no intrinsic structure, i.e., the number of contributing waves is too low to harness information on the interaction potential and collision dynamics from state-to-state experiments. 

As the energy is increased from the ultracold limit, the number of contributing partial waves consecutively increases yielding unprecendented opportunities to directly observe the effect of individual waves on the scattering cross section. Here, we enter the regime where scattering resonances occur in the collision cross sections: when the collision energy becomes resonant with a quasi-bound state supported by the interaction potential, the incident particles are temporarily captured into a quasi-bound long-lived complex \cite{Bell:MolPhys107:99}. In a simplified picture, these resonances may be regarded as the orbiting of the atom around the molecule (a shape resonance), or as the transient excitation of the molecule to a state of higher energy (Feshbach resonance). Analogous to Bohr’s model of electrons orbiting around an atom’s nucleus with given angular momentum, the atom-molecule binary system is stabilized when the collision energy is resonant with a partial wave that fits on the atom’s orbit an integer times. The corresponding "resonant" partial wave will dominate the scattering process, which appears as a sudden increase in the integral cross section (ICS) at the resonance energy. Most importantly, the resonant wave will dominate the differential cross section (DCS) as well, resulting in an angular distribution that fits the value for $\ell$ from which the resonance originates. This yields the unique opportunity to experimentally probe the partial wave fingerprint of the collision process, thereby revealing the scattering mechanism at the most fundamental state-to-state quantum mechanical level. 

Scattering resonances are extremely sensitive to the details of the PES, in particular at energies just above the ultracold Wigner limit where only a few waves with $\ell>0$ start contributing. Their observation has been a quest in molecular physics for decades \cite{Erlewein:Zphysik211:35}, however, it has proven extremely challenging to experimentally reach the low energies to access the relevant energy region and the high energy resolutions required to scan the energy over the resonances and to probe DCSs \cite{Naulin:IRPC33:427,Dong:Science327:1501,Yang:science347:60,Kim:Science349:510,Hauser:NatPhys6:467}. In ultracold gases, scattering resonances are routinely induced by tuning external fields to shift bound states into resonance \cite{RevModPhys.71.1,Quemener:CR112:4949}, but $s$-wave collisions observed here invariably result in isotropic DCSs, containing little information about the collision dynamics. To date, only a few experiments have succeeded in directly measuring the energy dependence of state-to-state cross sections near resonances, mostly using molecular beam methods \cite{Jankunas:ARPC66:241}. Resonances in ICSs have been observed using the merged beam \cite{Henson:Science338:234,Lavert-Ofir:NatChem6:332,Klein:NatPhys13:35,Shagam2015,Jankunas:JCP142:164305} and crossed beam approaches \cite{Chefdeville:PRL109:023201,Lara:PRL109:133201,Chefdeville:Science341:06092013,Bergeat:NatChem7:349, Bergeat:NatChem10:519}. Measurements of resonances features in state-to-state DCSs for inelastic collisions have recently also become possible using the Stark deceleration and velocity map imaging techniques \cite{Vogels:SCIENCE350:787,Vogels:NatChem10:435}, although the lowest energy achieved was limited to 13 cm$^{-1}$ which is typically two to three orders of magnitude too hot to probe individual partial waves. Measurements of state-to-state ICSs and DCSs at energies that approach the ultracold Wigner regime where only a few waves contribute have remained elusive, hampering a detailed view on how molecular collisions evolve from the pure quantum single-partial-wave into the semiclassical multi-partial-wave regime \cite{Kondov:PRL121:143401}. 

Here, we report the measurement of state-to-state ICSs and DCSs for inelastic collisions between state-selected NO ($X\,^2\Pi_{1/2}$, $v = 0,j = 1/2f$, hereafter referred to as $(1/2f)$ \cite{Note:1}) radicals and He atoms in a crossed beam experiment at energies between 0.2 and 8.5 cm$^{-1}$, with an energy resolution of 0.02 cm$^{-1}$. Three fully resolved partial wave resonances were observed in the ICS, and the incline of a fourth resonance at the lowest energies. These resonances originate from the lowest lying quasi-bound states supported by the NO-He interaction potential just above the energetic long-range asymptote. They correspond to values for $\ell$ that follow a progression from the lowest possible value, and probe the onset of the resonance regime at energies connecting to the ultracold Wigner limit. Rapid variations were observed in the highly structured DCSs as the energy was scanned over the resonances, where the number of nodes observed in the angular distributions directly reflected the increasing value of $\ell$ as the energy was increased. We found that our measurements of the lowest partial wave resonances in NO-He could not be reproduced satisfactorily by the most advanced NO-He PES at the CCSD(T) level of theory. Good agreement was only obtained with a new \emph{ab initio} PES at the CCSDT(Q) level. Such level of precision was previously only needed for high-resolution spectroscopy of bound states probing the PES in the region of the well \cite{Jankowski:Science336:1147}, and anticipated for low-energy scattering resonances \cite{Vogels:NatChem10:435}. The need for theory at the CCSDT(Q) level as demonstrated here for NO-He collisions illustrates the unprecedented sensitivity of scattering resonances in probing PESs -- accross their entire energy landscapes.

We used a crossed molecular beam apparatus that combines Stark deceleration and velocity map imaging (VMI), see Supplementary Information. A packet of NO ($1/2f$) radicals with a computer-controlled velocity and a narrow velocity spread was produced using the Stark decelerator. A beam of He atoms  was produced by a cryogenic valve held at temperatures between 11 and 18 K. Beam intersection angles of 5$^{\circ}$ and 10$^{\circ}$ were used to cover the 0.2 - 3 cm$^{-1}$ and 0.8 - 8.5 cm$^{-1}$ energy ranges, respectively. The collision energy resolution ranged from 0.02 cm$^{-1}$ for the lowest collision energies to 0.7 cm$^{-1}$ for the highest energies. The scattered NO radicals were state-selectively detected using a two-color laser ionization scheme and velocity mapped on a two dimensional detector. The measurements with a 5$^{\circ}$ intersection angle were performed using a VMI detector that offered improved velocity resolution. 

We studied collisions that de-excite the NO radicals from the ($1/2f$) to the ($1/2e$) level, which have an energy splitting of 0.01 cm$^{-1}$ \cite{meerts1972hyperfine}. For ICS measurements, we scanned the collision energy between 0.2 and 8.5 cm$^{-1}$, and observed three prominent resonances in the relative ICS with almost zero scattering probability at energies in between the resonances (see Fig. 1A). The incline of a fourth resonance was observed below 1 cm$^{-1}$. We compared the experimentally obtained relative ICS with theoretical ICSs, convoluted with the experimental resolution, based on two sets of PESs (See Fig. 1A and Supplementary Information). The first set of potentials was computed at the UCCSD(T)/CBS limit employing a complete basis-set extrapolation (CBS) scheme. The second set of PESs were computed at the UCCSDT(Q)/CBS level. We refer to the sets as CCSD(T)- and CCSDT(Q)-potentials, respectively. The CCSDT(Q)-potential is deeper than the CCSD(T) potential by about 0.6 cm$^{-1}$ near the well of the potential (see Supplementary Information), and causes a linear shift of roughly 0.25 cm$^{-1}$ in the position of all the calculated resonances. We found that the CCSDT(Q) corrections were essential to accurately describe the lowest-lying resonances, although the resonance observed near 6.5 cm$^{-1}$ was predicted at slightly lower energies by both potentials. A few test points computed at the UCCSDTQ(P) level suggest that the electron correlation for our UCCSDT(Q) potentials is converged to better than 0.1 cm$^{-1}$. We also devised a new method to investigate the sensitivity of the resonances to changes in the PESs: using an $S$-matrix Kohn variational method \cite{zhang:89,groenenboom:93b}, we isolate a square integrable resonance contribution to the wave function, which allows us to compute the first order response to changes in the PES as a simple integral. We tested overall scaling, scaling of the correlation energy alone, the anisotropy of the PESs or a radial shift (see Supplementary Information). We found that in all cases, all resonances shift in the same direction by nearly the same amount of energy, implying that it is hard to adjust the position of the highest resonance into closer agreement with the experiment without affecting the lower-lying resonances, even if we admit much larger variations than the expected uncertainty of the potential. With uncertainties in the PESs virtually ruled out as a non-negligible source of error in the resonance positions, we performed three further checks: we investigated (i) the effect of including scalar relativistic effects, (ii) the effect of explicitly including NO zero-point vibrational motion, and (iii) the effect of the diagonal Born Oppenheimer correction (DBOC). This last test required us to modify the usual DBOC, since our two PESs scattering method already corrects for the dominant non-Born-Oppenheimer couplings. These three checks are discussed in detail in the Supplement, and all are found to have a negligible effect. 

We analyzed the partial wave composition of the resonances from the CCSDT(Q) potentials, and expressed them in terms of the total angular momentum with conserved quantum number $\mathcal{J}$, which forms from the coupling of the rotational state of NO with quantum number $j=0.5$ and the partial wave quantum numbers $\ell_{in}$ and $\ell_{out}$ that represent the relevant partial waves in the ingoing and outgoing channel, respectively, i.e., $\mathcal{J} = \ell_{in/out} \pm 0.5$ (see Fig. 1B). Each observed resonance feature was found to consist of a set of overlapping resonances, that we could characterize as mostly pure Feshbach resonances (see Supplementary Information) with a unique value for the resonant partial wave $\ell_{res}$ corresponding to the quasi-bound NO-He state from which the resonance originates. The resonance below 1 cm$^{-1}$ corresponds to the lowest possible value for $\mathcal{J}$ and $\ell_{res}$ ($\mathcal{J}=0.5; \ell_{res}=2$), corresponding to the lowest NO-He quasi-bound state located just above the energetic asymptote of the free NO radical and He atom. Resonances with $\ell_{res}=0$ or 1 do not exist, as these values correspond to real bound states supported by the potential well (see Supplementary Information). The $\ell_{res}=2$ resonance is further characterized by almost equal contributions of outgoing waves with $\ell_{out}=0$ ($s$-wave) and $\ell_{out}=1$ ($p$-wave). From the ICS calculated in an extended energy regime (see Fig. 2), this resonance was found to connect to the pure $s$-wave Wigner regime where the ICS is proportional to $1/\sqrt{E}$ . The series of resonance features found at higher energies follow the progression of $\ell_{res}=2,3,4,5,....$, and probe the consecutive series of quasi-bound NO-He states supported by the interaction potential. 

The underlying partial wave composition of a resonance is directly reflected in the energy dependence of the DCS. We first analyzed the inherent structure of the DCS by compiling a contour plot of the theoretically predicted DCS versus collision energy, see Fig. 1C. We found that at energies in between the resonances, the DCS is rather independent of the collision energy. By contrast, at a resonance, the DCS was observed to change rapidly, following the rapid change of contributing partial waves. At the resonance energies, the two dominant ingoing and outgoing partial waves $\ell_{in}$ and $\ell_{out}$ underlying the resonance can clearly be discerned from the DCS: the number of nodes observed in the angular distribution equals the highest value of $\ell$.         

We experimentally probed the energy dependence of the DCS by measuring scattering images at selected energies where the DCS was predicted to change rapidly with collision energy, and compared them to a simulated scattering image based on the CCSDT(Q) potentials and kinematics of the experiment (Fig. 3). At the lowest energies probed, the velocity images are very small, but we could still discern structure in the images. For energies up to 0.7 cm$^{-1}$, we found strong backscattering, which evolved into a forward-backward peaked structure at energies around 0.7-0.8 cm$^{-1}$. The observed distributions arise from the interference of the $s$ and $p$-waves that dominate the scattering at these low energies. For energies above 0.8 cm$^{-1}$, we observed a rich energy dependence of the DCS, with additional peaks and nodes appearing directly reflecting the addition of consecutive partial waves as the energy was increased. Angular scattering distributions were extracted from the experimental and simulated images, and excellent agreement was found with the experimental and simulated distributions.     

The resonance structures in both the ICS and DCS as reported here are extremely sensitive to the details of the PESs: the small change in well-depth between the CCSD(T) and CCSDT(Q) PESs causes a significant and experimentally observable shift in resonance position. We investigated this sensitivity further by measuring the resonance position found around 1.3 cm$^{-1}$ for collisions of He with $^{14}$NO and $^{15}$NO, see Fig. \ref{fig:ICS_1415}. The isotope effect and associated change in reduced mass of the NO-He complex result in a difference in the centrifugal barrier and energies of the quasi-bound states supported by the potential. Consequently, the resonance position for He collisions with the heavier $^{15}$NO radical was found to shift 
by approximately 0.18 cm$^{-1}$ to lower energies, in excellent agreement with theoretical predictions. 

Our joint experimental and theoretical studies of partial wave resonances for NO-He collisions at energies down to 0.2 cm$^{-1}$ show that cold collisions in chemically relevant systems can now be probed with full quantum state resolution -- at the partial wave level --, and at energies approaching the Wigner regime. Measurements of the collision energy dependence of the ICS and in particular the DCSs revealed how molecular collisions transform from ultacold into hot by subsequently adding a partial wave as the energy is increased, thereby bridging the gap between the ultracold quantum physics and physical chemistry communities. The resonances are extremely sensitive to the details of the interaction potential, and standard theoretical methods at the gold-standard CCSD(T) level are not sufficiently accurate to predict the scattering cross sections, even for benchmark systems such as NO-He. The energies attained here are lower than the typical interaction energy of a molecule that posseses an electric or magnetic dipole moment with an external electric or magnetic field, respectively. This opens the possibility to tune the collision dynamics with tailored electric or magnetic fields, maintaining the quantum state preparation of the reagents and the quantum state resolution of the products. This would transform these studies from merely \emph{probing} nature with the highest possible level of detail into \emph{manipulating} nature with the highest possible level of control.

\bibliographystyle{Science}

\begin{thebibliography}{10}
\bibitem{Note:1}
The labels $X^2\Pi_{1/2}, v$, and $j$ indicate the electronic state, the
  vibrational state and rotational state of the NO radical, respectively. Each
  rotational state of NO possesses two near-degenerate $\Lambda$-doublet
  components, with symmetry labels $e$ (lower component) and $f$ (upper
  component), that refer to the total parity of the electronic wavefunction,
  exclusive of rotation.
\end{thebibliography}

{\bf Acknowledgement} 
This work is part of the research program of the Netherlands Organization for Scientific Research (NWO). S.Y.T.v.d.M. acknowledges support from the European Research Council (ERC) under the European Union's Seventh Framework Program (FP7/2007-2013/ERC Grant Agreement No. 335646 MOLBIL) and from the ERC under the European Union's Horizon 2020 Research and Innovation Program (Grant Agreement No. 817947 FICOMOL). We thank Jacek K{\l}os for providing us with NO-He PESs at the CCSD(T) level, and Theo Cremers for developing data acquisition software. We thank Niek Janssen and Andr\'e van Roij for expert technical support.

\begin{figure}
	\centering 
	\includegraphics[width=0.8\linewidth]{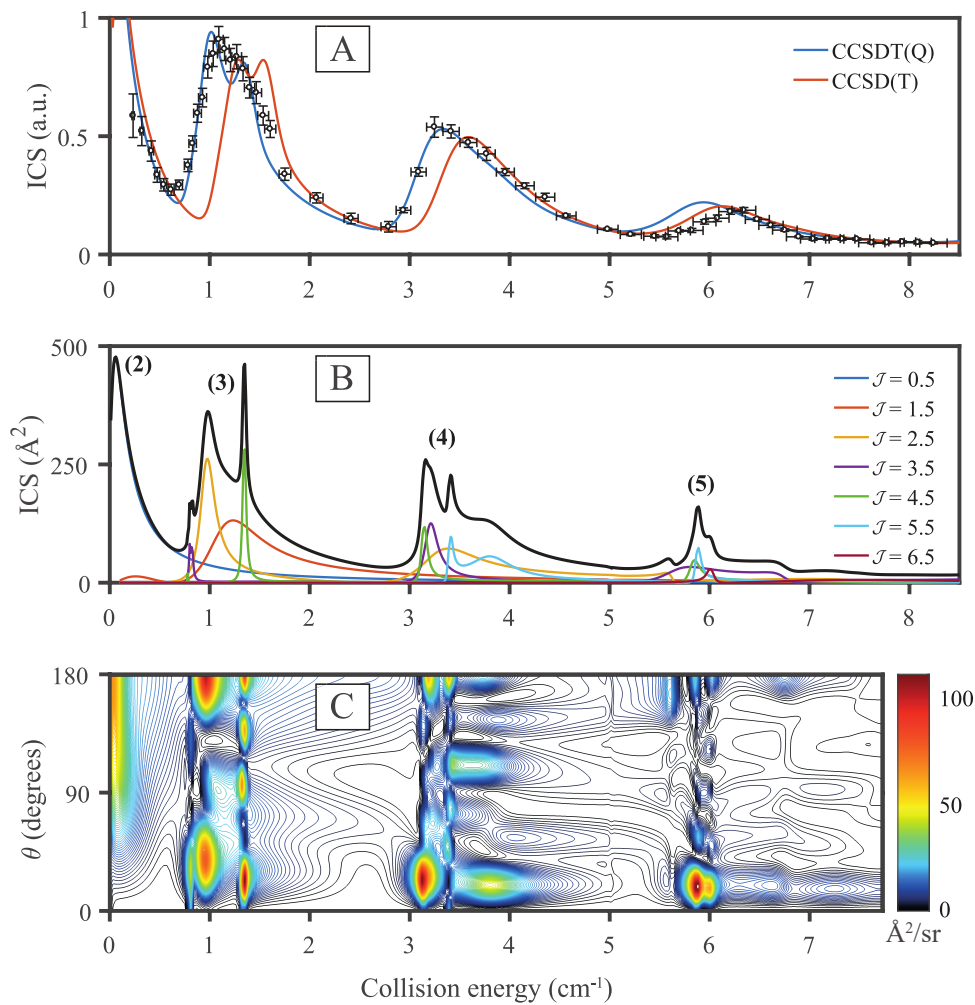}
	\caption{Collision energy dependence of integral and differential cross sections for inelastic NO-He ($j=0.5 f \rightarrow j=0.5e$) collisions. (A) Comparison between measured (data points with error bars) and calculated (solid curves) cross sections based on CCSD(T) and CCSDT(Q) potentials. Experimental data in a.u., arbitrary units. Data is accumulated using a continuous cycle over collision energies. Vertical error bars represent statistical uncertainties at 95\% of the confidence interval. Horizontal error bars represent uncertainties in energy calibration. The calculated cross sections were convoluted with the experimental energy resolution. (B) Theoretically predicted state-to-state cross section based on the CCSDT(Q) potential (black), together with the contribution of each angular momentum state with quantum number $\mathcal{J}$. The value for the resonant partial wave $\ell_{res}$ is given in parenthesis. (C) Contour plot of theoretically calculated (CCSDT(Q)-potential) DCS as function of collision energy. }	
	\label{fig:contour}
\end{figure}

\begin{figure}
	\centering
	\includegraphics[width=0.5\columnwidth]{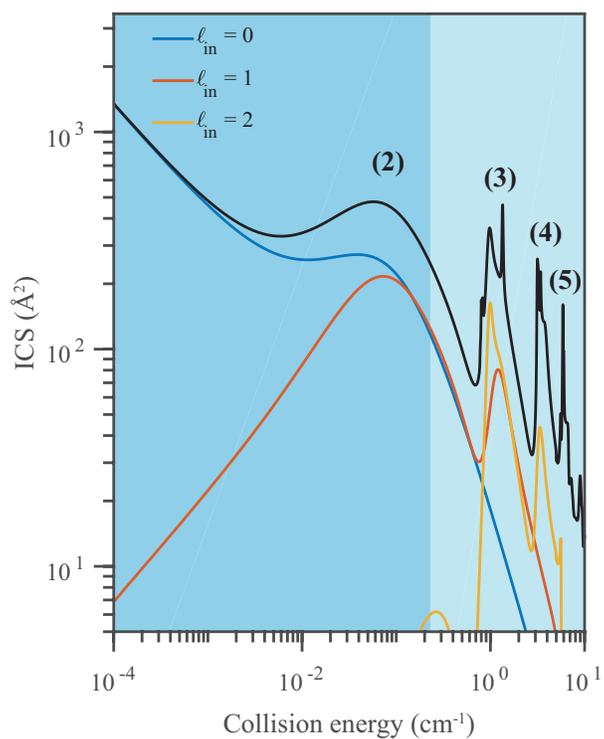}
	\caption{Logarithmic plot of theoretical ICSs based on the CCSDT(Q)-potential, including the contribution of $s$, $p$ and $d$ partial waves in the ingoing channel. The value for the resonant partial wave $\ell_{res}$ is given in parenthesis. The light blue shaded area indicates the energy region probed in our measurements. }
	\label{fig:ICS_log}
\end{figure}

\begin{figure}
	\centering
	\includegraphics[width=\columnwidth]{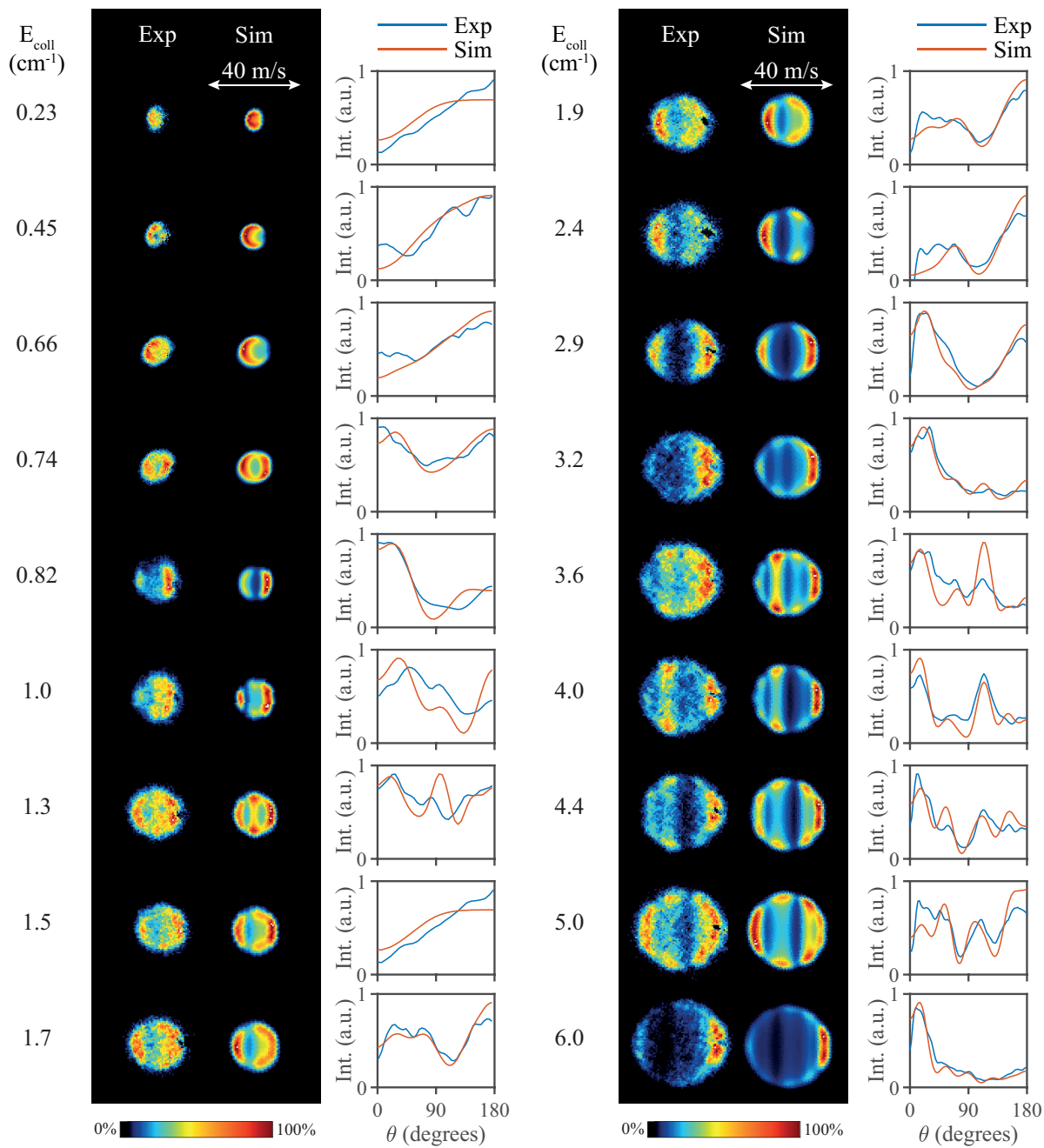}
	\caption{Experimental (Exp) and simulated (Sim) ion images based on the CCSDT(Q) potential at selected collision energies. The images are presented such that the relative velocity vector is oriented horizontally, with the forward direction on the right side of the image. Small
segments of the images around forward scattering are masked due to imperfect state selection of the NO packet. The angular scattering distributions as derived from the experimental (blue curves) and simulated (red curves) images are shown for each channel and collision energy. }
	\label{fig:DCS_M}
\end{figure}

\begin{figure}
	\centering
	\includegraphics[width=0.5\columnwidth]{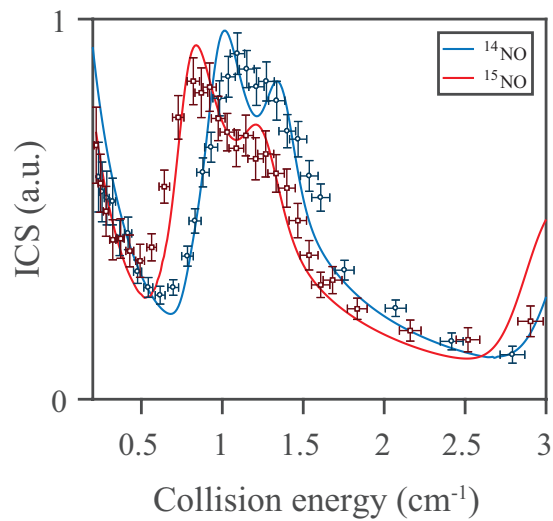}
	\caption{Measurements of the relative state-to-state ICSs for inelastic ($j=0.5 f \rightarrow j=0.5e$) collisions of $^{14}$N$^{16}$O (blue data points) and $^{15}$N$^{16}$O (red data points) with He around the resonance feature at 1 cm$^{-1}$, together with the calculated cross sections from the CCSDT(Q) potential convoluted with the experimental resolution. }
	\label{fig:ICS_1415}
\end{figure}

\end{document}